\documentclass[12pt,preprint]{aastex}

\tighten
\def\<<{{\ll}}
\def\>>{{\gg}}

\def\spose#1{\hbox to 0pt{#1\hss}}
\def\ltwig{\mathrel{\spose{\lower 3pt\hbox{$\mathchar"218$}}
     \raise 2.0pt\hbox{$\mathchar"13C$}}}
\def\gtwig{\mathrel{\spose{\lower 3pt\hbox{$\mathchar"218$}}
     \raise 2.0pt\hbox{$\mathchar"13E$}}}
\def\+/-{{\pm}}
\def\=={{\equiv}}
\def\Rstar{R_{\ast}}

\def\vth{v_{th}}

\def\Mdot{\dot M}
\def\mdot{\dot m}

\newcommand{\beq}{\begin{equation}}
\newcommand{\eeq}{\end{equation}}
\newcommand{\beqa}{\begin{eqnarray}}
\newcommand{\eeqa}{\end{eqnarray}}

\ifx\pdfoutput\undefined 
\usepackage{graphicx} 
\else 
\usepackage[pdftex]{graphicx} 
\usepackage{epstopdf} 
\fi 

\begin{document}
\title{The Effect of Magnetic Field Tilt and Divergence on the ~~~~~
Mass Flux and Flow Speed in a Line-Driven Stellar Wind}
\author{Stanley P. Owocki}
\affil{Bartol Research Institute,
       University of Delaware,
       Newark, DE 19716}
\author{Asif ud-Doula}
\affil{Department of Physics,
       North Carolina State University, Box 8202
       Raleigh, NC 27695-8202}   

\date{\today}
\begin{abstract}
We carry out an extended analytic study of how the tilt and 
faster-than-radial expansion from a  magnetic field affect 
the mass flux and flow speed of a line-driven stellar wind.
A key motivation is to reconcile results of numerical MHD simulations 
with previous analyses that had predicted non-spherical expansion 
would lead to a strong speed enhancement.
By including finite-disk correction effects, a dynamically more 
consistent form for the non-spherical expansion, and a moderate value 
of the line-driving power index $\alpha$, we infer more modest speed 
enhancements that are in good quantitative agreement with MHD simulations, 
and also are more consistent with observational results.
Our analysis also explains simulation results that show the 
latitudinal variation of the surface mass flux scales with the square of 
the cosine of the local tilt angle between the magnetic field 
and the radial direction.
Finally, we present a perturbation analysis of the effects of a 
finite gas pressure on the wind mass loss rate and flow speed in both 
spherical and magnetic wind models, showing that these scale with the 
ratio of the sound speed to surface escape speed, $a/v_{esc}$, and are 
typically 10-20\% compared to an idealized, zero-gas-pressure model.
\end{abstract}

\section{Introduction}
In a recent paper (ud-Doula \& Owocki 2002; hereafter UO-02), we  
presented
numerical magnetohydrodynamic (MHD) simulations of the effect of a
stellar dipole magnetic field on the line-driven stellar wind from a 
non-rotating,
hot, luminous star. We showed that the overall
influence of the field on the wind depends largely on a
single, dimensionless, `wind magnetic confinement parameter',
$\eta_{\ast}$ ($ = B_{eq}^2 R_\ast^2/{\dot M} v_\infty$), which
characterizes the ratio between magnetic field energy density and
wind kinetic energy density.
Because the field energy declines faster than the
wind energy,  in the outer regions the magnetic field
is  always dominated by the radial wind outflow,
which thus asymptotically stretches the field into a radial 
configuration, regardless of the strength of $\eta_*$.
For weak confinement $\eta_{\ast} < 1$, this radial opening of the field 
applies throughout the whole computational domain. 
But for stronger confinement $\eta_{\ast} > 1$, the magnetic field remains 
closed near the surface and over a limited range of latitude around the
magnetic equator.

In this paper we provide further analysis and interpretation
of how key wind properties, namely the 
surface mass flux and asymptotic flow speed 
(see figs. \ref{fig1}a,b in \S 2 below), 
are modified by the presence of a magnetic field.
One central motivation is to reconcile results of
our numerical MHD simulations with the previous scaling analysis done
by MacGregor (1988; hereafter M-88), which had predicted that the
faster-than-radial divergence of magnetic flux tubes would lead to substantial 
(factor 3-4) increase in the terminal velocity of a line-driven wind, compared to 
that of the non-magnetic, spherical-expansion case.
Since the implied flow speeds ranging up to 5000-6000 km/s are
never observed through blue edges of  UV P Cygni lines from hot stars
(Prinja \& Howarth 1986; Howarth and Prinja 1987), that prediction could be a basis
for inferring that hot stars lack  magnetic fields with sufficient 
strength to substantially influence their stellar winds.

Here we develop (\S 3) a simplified formulation of the 
one-dimensional (1D) equations for a steady line-driven wind.
Applying this to cases with faster-than-radial area expansion (\S 4),
we show that the prediction of a 3-fold  increase in flow speed 
is the consequence of certain assumptions and idealizations
(point-star radiation, phenomenological flow expansion, and  large 
line 
power-index $\alpha$) in the previous M-88 analysis;
when these assumptions are relaxed (to include finite-disk 
correction, a dynamical flow divergence, and more realistic $\alpha$)
the expected effect on flow speeds is consistent with the more modest 
$ \ltwig 50 \%$ increase typically found in our MHD simulations.
We further show that the inferred simulation
scaling of the base mass flux with the square of the
cosine of the surface field tilt angle can be understood
from a simple modified 1D tilted flow analysis (\S 5).
After a brief discussion of the implications for stellar wind structure (\S 6), 
we conclude with a summary (\S 7) of our main results.

\section{Summary of Key Results of 2D MHD Simulations}

\begin{figure}
\begin{center}
\plotone{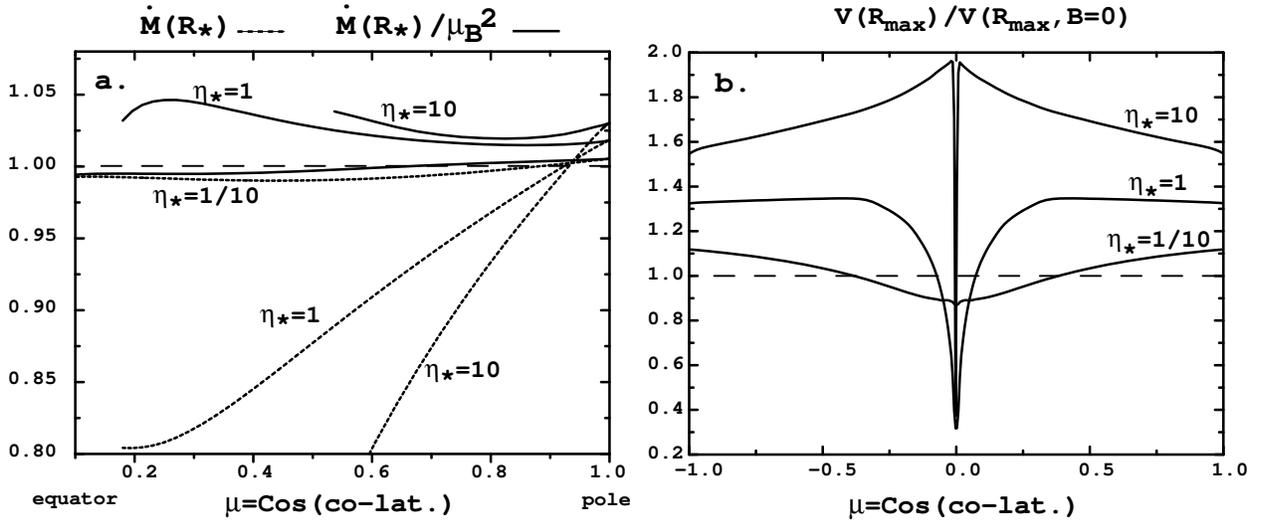}
\caption{
a. The latitudinal variation of the base mass flux in open field 
regions of 2D MHD simulations.
The dotted curves represent the mass flux scaled by the corresponding 
spherical model.
The solid curves add an additional scaling 
by $\mu_{B}^{2}$, where $\mu_{B} \equiv {\hat B_{\ast}} \cdot {\hat r}$ is the 
radial projection of a unit vector along the base magnetic field.
b. Latitudinal variation of the flow speed at the maximum radius 
$R_{max} =  6 \Rstar$ of these same 2D MHD simulations, scaled by the 
corresponding speed in the spherical, non-magnetic ($B=0$) model.
}
\vspace{-0.2cm}
\label{fig1}
\end{center}
\end{figure}

Figures \ref{fig1}a,b summarize
two key results of our 2D MHD simulations,
wherein we study the dynamical competition between magnetic field and wind 
by inserting a dipole magnetic field of varied strength 
(parameterized by $\eta_{\ast}$)  
into a previously relaxed spherically symmetric CAK wind.\footnote{
The simulation results given here are analogous to those presented 
by UO-02 and ud-Doula (2002), but correct a minor coding error that 
lead the polar flow to have an artifically enhanced mass flux and an 
artificially reduced flow speed.
As in UO-02, the model here is assumed isothermal, with a constant sound 
speed $a=26.3$~km/s, implying a ratio $s \equiv a^{2}/v_{esc}^{2} = 0.0014$ 
between the gas internal energy and surface escape energy. (See \S 3 
and the Appendices.)
}
For the asymptotic states when the regions of magnetic field opened by the 
wind have a nearly steady-state flow, 
both figures show results for three cases, with the degree of 
magnetic confinement ranging from weak ($\eta_{\ast}=0.1$), through 
moderate ($\eta_{\ast}=1$), to strong ($\eta_{\ast}=10$).

In fig.~\ref{fig1}a, the dotted curves show the latitudinal variation 
of the base mass flux, while the solid curves show this same flux scaled 
by $\mu_{B}^{2}$, where $\mu_{B}$ is the radial projection of a unit 
vector along the base magnetic field.
The near constancy of this scaled flux is explained in \S 5, which uses a 
simple tilted-flow analysis of the base line-driving to show that the 
radial mass flux does indeed vary with the square of the field's radial 
projection cosine, ${\dot M} \sim \mu_{B}^{2}$.

At the magnetic poles, the radial orientation of the field  means the 
flow is also radial, with essentially the same base mass flux as in 
the spherical model.\footnote{
The few percent increase is a second-order effect, associated with 
the finite sound speed and the resulting detachment of the flow 
critical point from the surface, where faster areal expansion reduces 
the density, and so allows driving of a somewhat greater mass flux. 
See Appendix B.
}

Fig.~\ref{fig1}b plots the latitudinal variation of the flow speed 
at the maximum model radius $R_{max} = 6 \Rstar$, scaled by the 
corresponding speed in the spherical, non-magnetic ($B=0$) model.
The lower speeds toward the magnetic equator ($\mu \approx 0$) reflect 
the various relative degrees of equatorial flow compression, which 
increases the density, and so reduces the radiative acceleration.
In the highest  magnetic confinement case $\eta_{\ast} = 10$, the 
increased speeds at mid-latitudes ($0.1 < |\mu| < 0.8$)  result from 
a combination of the reduced mass flux and greater flow expansion, both of 
which reduce the density, and so increase the radiative acceleration.

While such latitudinal trends thus make sense qualitatively, our focus 
below will be on developing a semi-quantitative understanding of the 
wind properties at the magnetic poles, where the  radial orientation
of the flow makes possible a 1D analysis, if one takes proper account
of the non-spherical expansion of the flow area.
This was essentially the approach taken by M-88, but the factor 3 or 
more speed enhancements found there are substantially greater than 
those in fig.~\ref{fig1}b, which even for the strongest magnetic 
confinement $\eta_{\ast} = 10$ shows a polar speed enhancement of only  
$1.54$.
The next two sections develop and apply a formalism for understanding 
this difference.

\section{Flow Analyses for Spherical Expansion}

\subsection{CAK Model of a Wind Driven by a Point-Source of Radiation}

To provide a basis for our analysis of how the tilt and areal 
divergence of a magnetic field modify a wind outflow, let us first 
recast the basic Castor, Abbott and Klein (1975; hereafter CAK) formalism 
in a somewhat more tractable, physically transparent form.
First, for a steady, !D stellar wind driven radially by line-scattering of 
radiation from a central point-source, the associated equation of motion for 
the radial velocity $v$ as a function of radius $r$ takes the form
\beq
\left [ 1 - {a^{2} \over v^{2} } \right ] \, v {dv \over dr} 
=- \frac {GM(1-\Gamma)}{r^2} \; + \; g_{CAK} 
 + { 2 a^{2} \over r }
\, ,
\label {eom}
\eeq
where $GM(1-\Gamma)/r^2$ is the effective gravitational acceleration,
and the Eddington parameter $\Gamma \equiv\kappa_e L/4 \pi GMc$ 
accounts for the effect of electron scattering opacity $\kappa_e$ 
interacting with the stellar luminosity $L$, with $c$ the speed oflight.
The square-bracket factor with the sound speed $a$ on the left-hand-side 
accounts for gas pressure effects that allow for smooth mapping of the wind 
outflow onto a hydrostatic atmosphere below the sonic point, where $v=a$.
On the right-hand-side the flow's areal expansion also gives rise to a 
pressure-form term, $2a^{2}/r$, which compared to the competing gravitational 
term is of order $s \equiv (a/v_{esc})^{2} \approx 0.001$, where 
$v_{esc} \equiv \sqrt{2GM(1-\Gamma)/R_{\ast}}$ is the effective 
escape speed from the stellar surface radius $R_{\ast}$. 
The outward driving against gravity through the sonic point and 
beyond must thus be principally via the line acceleration, which under CAK's 
original assumption of radially streaming radiation from a point
source can be written as
\beq
\label{gcak}
g_{CAK}
=\frac{1}{(1-\alpha)} ~ \frac{\kappa_e L{\bar {Q}}}{4 \pi r^2 c }
\left (\frac{dv/dr}{\rho c \bar Q \kappa_e}\right )^{\alpha}
\eeq
where $\rho$ is  the density, and $\alpha$ is the CAK exponent.
Here the line normalization $\bar Q$ is related to the usual CAK 
``k'' parameter through
$k= {\bar Q}^{1-\alpha} \, \left ( \vth/c \right)^{\alpha}/(1-\alpha)$,
but offers the advantages of being a dimensionless
measure of line opacity that is independent of the assumed ion thermal
speed $\vth$,  with a nearly constant characteristic value of order
${\bar Q} \sim 10^3$ for a wide range of ionization conditions 
(Gayley 1995).

To proceed let us define the gravitationally scaled inertial acceleration
\beq
w' \equiv \frac
{r^2 v dv/dr}{GM (1-\Gamma)}\, .
\label{wp-def}
\eeq
In terms of an inverse radius coordinate $x \equiv 1-R_{\ast}/r$,
note that $w' = dw/dx$, where $w \equiv v^{2}/v_{esc}^{2}$ represents 
the ratio of wind kinetic energy to the effective gravitational binding 
$v_{esc}^{2}/2 \equiv GM(1-\Gamma)/R$ 
from the stellar surface radius $R_{\ast}$.
We can then rewrite the equation of motion (\ref{eom})
in the dimensionless form
\beq
(1-s/w) \, w' = -1 + C (w')^\alpha \, + 2s/(1-x)
\label{cakeom}
\eeq
where $s \equiv a^{2}/v_{esc}^{2}$, and 
we have eliminated the density $\rho$ in favor of the mass loss rate
$\Mdot=4 \pi r^2 \rho v$ for the assumed steady, spherical expansion,
with the line-force constant defined by
\beq
C \equiv
\frac {1}{1-\alpha} \, \left [ \frac {L} {\Mdot c^2} \right ]^\alpha 
\,
\left [ \frac {{\bar Q} \Gamma} {1-\Gamma} \right ]^{1-\alpha} \, .
\label{cdef}
\eeq
Note that, for fixed sets of parameters for the star ($L$, $M$, $\Gamma$) 
and line-opacity ($\alpha$, $\bar Q$), this constant scales with the mass 
loss rate as $C \sim 1/\Mdot^{\alpha}$.

\begin{figure}
\begin{center}
\plotone{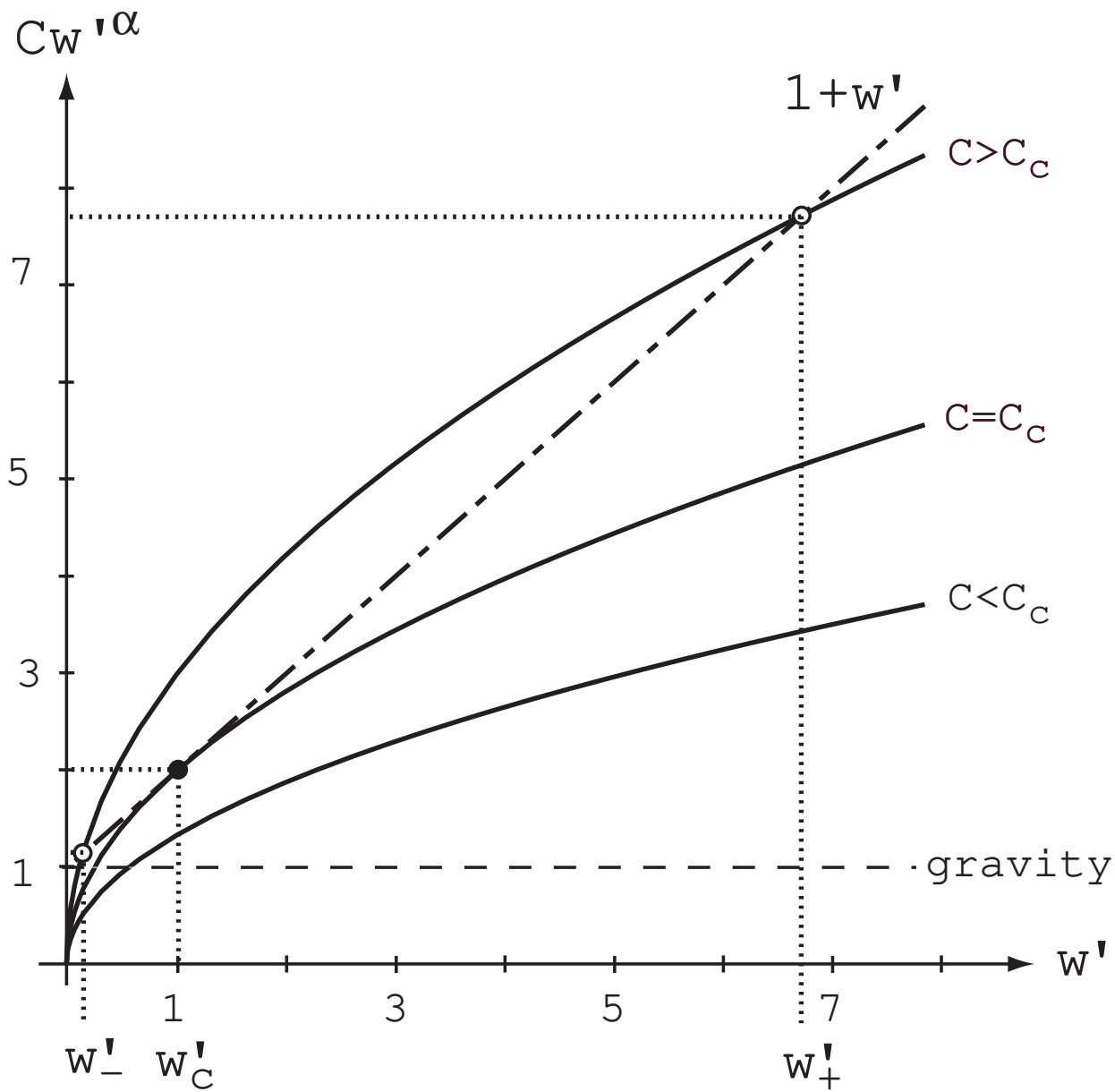}
\caption [1D-CAK Wind Solutions]{
Graphical solution of the dimensionless equation
of motion (\ref{cakeom}) representing a 1D, point-star, 
zero-sound-speed CAK wind, as controlled by the constant 
$C \sim 1/\Mdot^{\alpha}$, with determine the intersections of the 
line-force $C w'^{\alpha}$ (solid lines) with the combined gravity 
plus inertia $1+w'$ (dot-dashed lines).
If  $\Mdot$ is too big, there are no solutions;
if $\Mdot$ is small there are two solutions.
A maximal value $\Mdot= \Mdot_{CAK}$ defines a single, ``critical'' solution.
}
\label{fig2}
\end{center}
\end{figure}

The essential elements of interest in this paper can be analyzed 
without including the small gas pressure terms, and so we shall 
henceforth consider only  the limit of vanishing sound speed 
$a \sim \sqrt{s} \rightarrow 0$.
(The corrections from a small but finite $s$ are examined in the 
Appendices.)
For this limit, fig.~\ref{fig2} illustrates a graphical solution to eqn. 
(\ref{cakeom}) for various values of $C$.
For high $\Mdot$ or small $C$ there are no solutions,
while for small $\Mdot$ or high $C$ there are two solutions.
The CAK critical solution corresponds to a {\it maximal} mass loss 
rate, and requires a {\it tangential} intersection between line-force and
combined inertia plus gravity, for which
\beq
\alpha \, C_c  {w'_c}^{\alpha-1}=1 \, ,
\label {wder}
\eeq
and thus
\beq
w'_c= \frac {\alpha}  {1-\alpha} \, ,
\label {weq}
\eeq
with
\beq
C_c= \frac {1} {\alpha^\alpha (1-\alpha)^{1 - \alpha}} \, .
\label{c-def}
\eeq
Applying this in eqn. (\ref{cdef}), we can then solve for the mass loss 
rate, yielding the standard CAK scaling 
\beq
\Mdot_{CAK}=\frac {L}{c^2} \; \frac {\alpha}{1-\alpha} \;
{\left[ \frac {\bar {Q} \Gamma}{1- \Gamma} 
\right]}^{{(1-\alpha)}/{\alpha}} \, .
\label{mdot}
\eeq
Moreover, since the scaled equation of motion (\ref{cakeom}) has no 
explicit spatial  dependence, the scaled critical acceleration $w'_c$
applies throughout the wind.
This can thus be trivially integrated to yield 
$w = \alpha x/(1-\alpha )$, which is equivalent to the CAK velocity law
\beq
v(r)=
v_\infty
\left ( 1- \frac {R_{\ast}}{r} \right )^{1/2} \, ,
\label{CAK-vlaw}
\eeq
where the wind terminal speed $v_\infty = v_{esc} 
\sqrt{\alpha/(1-\alpha)}$ scales with
$v_{esc} \equiv \sqrt{2GM(1-\Gamma)/R_{\ast}}$, the effective escape 
speed from the stellar surface radius $R_{\ast}$.

\subsection{The Finite-Disk Correction Factor}

The above analysis has so far been based on the idealization of 
radially streaming radiation, as if the star were a point-source 
at the origin.
This was the basis of the original CAK wind solutions, although they 
did already identify (but did not implement) the appropriate 
``finite-disk correction factor'' to account for
the full angular extent of the star [See CAK eqn. (50).],
\beq
f_d=\frac {(1+\sigma)^{1+\alpha}-(1+ \sigma \mu_\ast^2 )^{1+\alpha}}
{(1+\alpha) \sigma (1+\sigma)^\alpha (1-\mu_\ast^2)} \, ,
\label{fd-def}
\eeq
where $\sigma \equiv (r/v)dv/dr  \, -1$ 
and $\mu_{\ast}^{2} \equiv 1-R_{\ast}^{2}/r^{2}$.
When this factor is included to modify the point-star CAK 
line-acceleration (\ref{gcak}), its complex dependence on radius, 
velocity, and velocity gradient complicates somewhat the solution 
of  the full equation of motion (\ref{eom}).
But full solutions have been derived independently by 
Friend and Abbott (1986) and Pauldrach, Puls, and Kudritzki (1986).
They yield a somewhat reduced mass loss rate 
$\Mdot_{fd} \approx \Mdot_{CAK} / (1+\alpha)^{1/\alpha}$,
higher terminal speed $ v_\infty \approx 3 v_{esc}$,
and flatter velocity law, roughly characterized by replacing
the exponent $1/2$ in eqn.  (\ref{CAK-vlaw}) with a somewhat higher 
value $\beta \approx 0.8$.

\begin{figure}
\begin{center}
\plotone{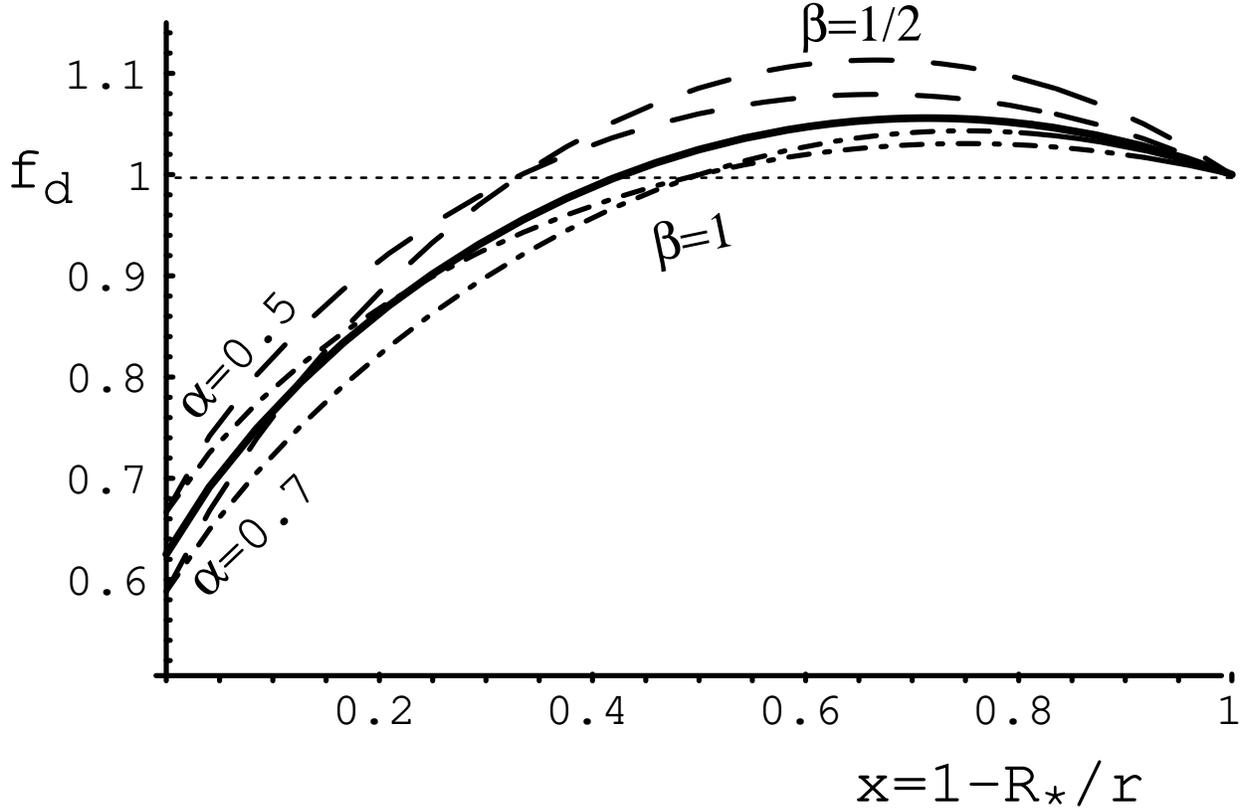}
\caption{Spatial variation of the finite-disk-correction factor 
$f_{d}$, plotted vs. scaled inverse radius $x = 1 - R_{\ast}/r$,
for various  CAK exponents $\alpha$ and velocity-law exponents $\beta$.
The heavy solid curve shows our standard case with $\alpha=0.6$ and 
$\beta=3/4$.
The dashed curves use $\beta=1/2$ and the dot-dashed curves 
$\beta=1$, and are each plotted for CAK exponents $\alpha=0.5$ and 
$\alpha=0.7$ that span the nominal range.
A key point is that all these varied curves have a similar form, 
rising sharply from a base value [$f_{d\ast}=1/(1+\alpha)$], 
crossing above unity (dotted line), 
and then relaxing back to it at large radial distances, 
$x \rightarrow 1$.}
\vspace{-0.2cm}
\label{fig3}
\end{center}
\end{figure}

If one simply assumes such a canonical ``beta'' velocity law, 
fig.~(\ref{fig3}) illustrates
the specific spatial variations of $f_d(x)$ for various CAK exponents 
$\alpha$
and velocity exponents $\beta$.
All these varied cases show a quite common general behavior.
For the initial acceleration from a static stellar surface 
($r \approx R_\ast$, or $x \rightarrow 0$; where $v \rightarrow 0$ with finite 
$dv/dr$, implying $\sigma  \rightarrow \infty$),
$f_d$ has a starting value $f_{d\ast} \approx 1/(1+\alpha)$.
It then increases outward, past unity at the point where the
wind has a locally isotropic expansion (with $dv/dr=v/r$, so that 
$\sigma=0$), and then finally falls back asymptotically toward unity 
at large radii ($x \rightarrow 1$), 
where the star is indeed well approximated by a point source.

To understand the dynamical effect of this finite-disk factor, and 
indeed to see how it interacts with effects of a magnetic field, 
it is helpful to examine how such a typical radial variation influences 
the scaled acceleration $w'$ through solutions to the appropriately 
modified equation of motion [cf. eqn (\ref {cakeom})],
\beq
w' = - 1 + f \, C_* \,  (w')^\alpha \,
,
\label {fdeom}
\eeq
where for convenience we have defined scaled quantities 
$f \equiv f_d /f_{d\ast}$ and $C_\ast \equiv f_{d \ast} C$,
and, for simplicity, again taken the zero-sound-speed limit $s=0$.
Since $f$ increases outward from unity at the surface, the surface 
radius $r=R_\ast$ ($x=0$)
now represents a critical point (also known as a choke or throat) 
that fixes 
the required
minimal value of $C_\ast = C_c = 1/\alpha^{\alpha} 
(1-\alpha)^{1-\alpha}$ to allow an
accelerating solution from this wind base.
Accounting then for the mass loss rate scaling of $C_\ast \sim 
1/\Mdot^\alpha$, we
readily see that the net result is to reduce the maximal allowed mass 
loss to
\beq
\Mdot_{fd} =
f_{d\ast}^{1/\alpha} \Mdot_{CAK} = \Mdot_{CAK}/(1+\alpha)^{1/\alpha} 
\, .
\label {mdfd}
\eeq
This thus provides a simple rationale for this aspect of the detailed 
numerical solutions of Friend and Abbott (1986) and Pauldrach et al.  (1986).

As the finite-disk factor increases from its reduced base value, the 
lower mass loading allows for stronger CAK line-force, with now the 
resulting greater acceleration being ``leveraged'' by its effect in 
further increasing the line-force.
From the graphical solution in fig.~\ref{fig2}, one sees 
that a modest increase in the line-acceleration term [$C (w')^\alpha$] 
above the critical solution (with $C=C_c$) leads to a solution with a 
{\it much} larger scaled acceleration $w'$.
[There also arises a much shallower acceleration, but as discussed by
Feldmeier et al. (2002), maintaining this beyond the critical 
point requires a special kind of outer boundary condition, whereas 
the more rapid acceleration is compatible with the standard assumption of 
simple expansion into a vacuum.]
\begin{figure}
\begin{center}
\plotone{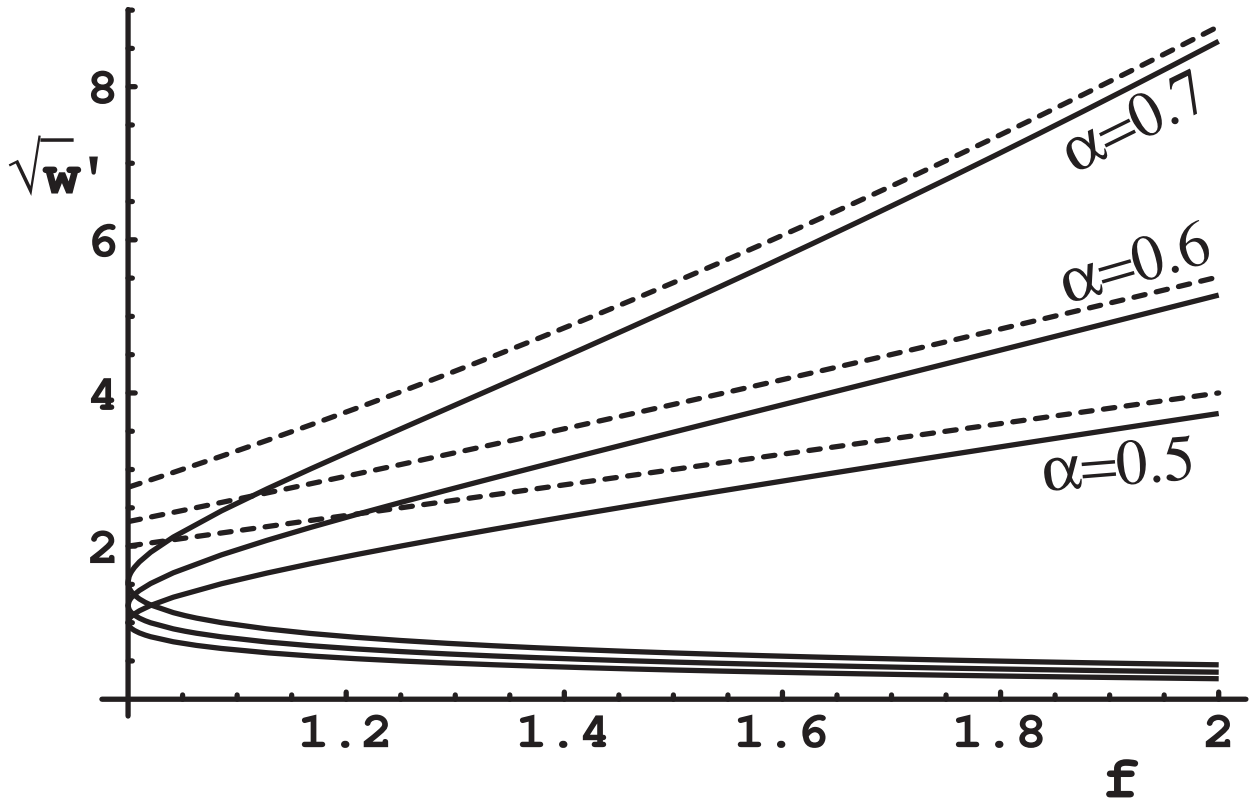}
\caption{
Root of the scaled acceleration $\sqrt{w'}$ that solves the 
equation of motion (\ref{fdeom}), plotted as a function of the 
surface-normalized 
finite-disk factor $f$.
Since the wind terminal speed scales as $v_{\infty} \sim \sqrt{w'} 
v_{esc}$, note that even modest increases in the finite-disk factor 
can lead to a very high speed wind (especially for larger $\alpha$).
The dashed curves denote the asymptotic solutions 
$\sim (C_{c}f)^{1/(2(1-\alpha))}$ that apply in the limit of large 
acceleration $w'\gg 1$.
}
\vspace{-0.2cm}
\label{fig4}
\end{center}
\end{figure}

Fig.~\ref{fig4} illustrates this strong sensitivity to the 
finite-disk correction in terms of $\sqrt{w'}$, which sets the 
scaling of terminal speed $v_\infty \sim v_{esc} \sqrt{w'}$.
Note how a modest, less-than-factor-two change in $f$ leads to large 
increases in this speed scale $\sqrt{w'}$, with the trends being most 
dramatic for the largest $\alpha$.
From eqn. (\ref{fdeom}), we see that in the limit of large 
acceleration $w' \gg 1$, the speed scaling approaches the form
\beq
\sqrt{w'} \sim \left ( C_{c} f \right )^{1/2(1-\alpha)} ~~ ; ~~ w' 
\gg 1 \, ,
\label{swplim}
\eeq
which is illustrated through the dashed curves in fig.~\ref{fig4}.
For the specific case $\alpha=1/2$, we can use the quadratic formula 
to obtain a simple expression that applies to the full range,
\beq
\sqrt{w'} = f \left [ 1 + \sqrt{1-1/f^2} \right ] \, ,
\label {wphalf}
\eeq
where the choice of the ``plus'' root gives the steeper of the two 
acceleration solutions.

\begin{figure}
\plotone{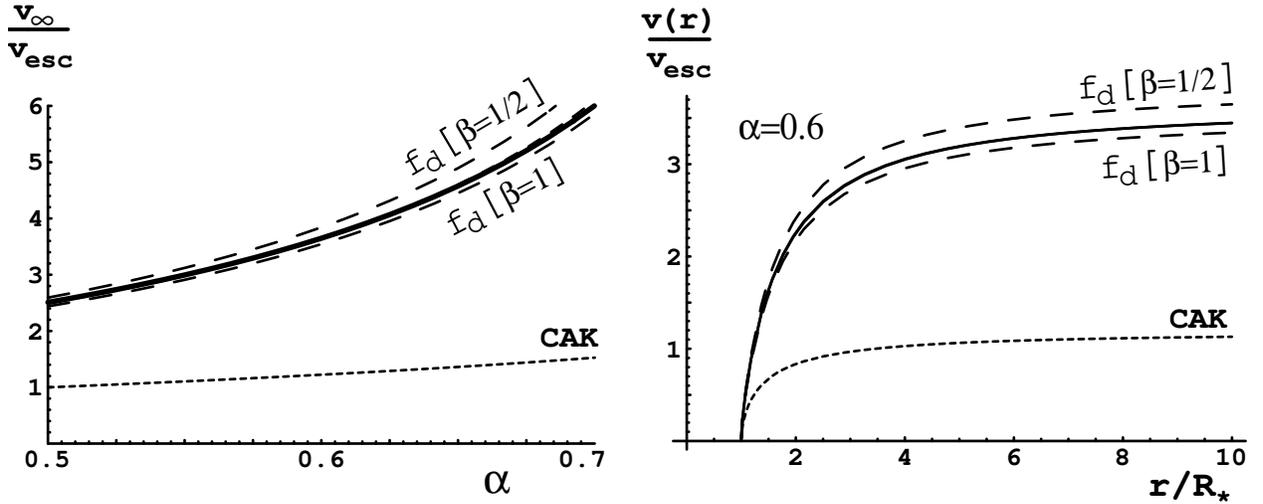}
\caption{
Comparison of the full finite-disk solutions (solid) with 
integrations that use a beta-velocity law to obtain spatially explicit 
approximations for the finite-disk factor (dashed).
Left: The ratio of terminal to escape speed vs. CAK exponent $\alpha$.
Right: Radial variation of the velocity for the $\alpha=0.6$ case.
In both panels, results for a model using $\beta=3/4$ to evaluate 
the finite-disk factor are also included, but are indistinguishable 
from the full solution plotted by the solid line.
The dotted curves show corresponding results for the point-star CAK 
model.
}
\vspace{-0.2cm}
\label{fig5}
\end{figure}

To obtain specific values for the flow speed, we can numerically 
integrate the scaled equation of motion (\ref{fdeom}) accounting for 
the complex functional dependence of the finite disk factor 
(\ref{fd-def}) on radius, velocity, and velocity gradient.
The left panel in fig.~\ref{fig5} shows how the ratio of terminal to 
escape speed depends on the CAK power index $\alpha$; the right 
panel plots the full radial variation of velocity for the $\alpha=0.6$ case.
The dashed curves compare results for a similar integration when a 
canonical beta velocity law is used to evaluate the finite-disk 
factor.
For $\beta=0.75$ the curves are indistinguishable from the full 
calculation, and other values of $\beta$ also give very good 
approximations.

A basic conclusion is thus that one may derive accurate 
finite-disk-corrected wind solutions using a beta velocity law to 
obtain spatially explicit approximations of the finite-disk factor, 
with especially good agreement obtained for $\beta \approx 0.75$.

\section{The Effect of Non-Spherical Area Expansion}

\subsection{
Kopp-Holzer Expansion in a Point-Star CAK Wind}

The above formulation of standard line-driven wind theory provides
a convenient basis for understanding the results of models extended to
include magnetic fields.
As first analyzed by MacGregor (M-88), one potentially key effect
regards the faster-than-radial expansion of the area in a flow tied
to a rapidly diverging magnetic field.
Following Kopp \& Holzer (1976), it is convenient
to consider an infinitesimal flow tube with a non-radial expansion, 
but still with a radial {\it orientation}, such as would apply, for 
example, to flow emanating from the poles of a surface magnetic dipole.
The non-radial expansion factor can then be written as
\beq
h(r) \equiv \frac {A (r) }{A_\ast}\,  \frac {R_{\ast}^2}{r^2}=
\frac {B (R_{\ast}) } {B(r)} \,
\frac {R_{\ast}^2} { r^2} \, ,
\label{h-def}
\eeq
where the latter equality follows from the 
$\nabla \cdot {\bf B} = 0 $ property of any
magnetic field, since that then implies $d(BA)/dr=0$ and so $A \sim 
1/B$ along such a radial flow tube.

In the absence of a detailed dynamical solution for the magnetic 
field divergence, M-88 adopted the Kopp \& Holzer (1976) phenomenological
non-spherical expansion factor\footnote{
We use the notation $h$ instead of Kopp and Holzer's $f$ to
distinguish from the finite-disk correction defined above.},
\beq
h_{KH}={ h_{max} \exp[(r-R_1)/\sigma]+ 1 - (h_{max}-1) 
\exp[(R_\ast-R_1)/\sigma]
    \over
    \exp[(r-R_1)/\sigma] + 1
    } \, ,
\label{fkhdef}
\eeq
where $R_1$, $\sigma$, and $h_{max}$ are free parameters, the first 
two specifying the location and extent of the non-spherical expansion region, 
and the last fixing the total overall level of non-spherical expansion.
Following M-88, we fix sample values of the two Kopp-Holzer parameters, 
$R_1=1.25 R_\ast$ and $\sigma=0.1$, and compare models with various 
$h_{max} =$~1.5 and 2.

Applying this Kopp \& Holzer expansion factor to CAK line-driven
wind models computed in  the idealized point-star approximation, M-88 
found substantial (more than factor three) increases in wind terminal speed 
for even modest values of the total non-spherical expansion, 
$h_{max} \approx 2$.
The above formalism provides a convenient way to understand those 
results, and indeed to extend them to account for the combined effects of a 
non-spherical expansion and a finite-disk correction. (See \S 4.2.)
Noting from eqn. (\ref{gcak}) that the line acceleration scales with 
the density $\rho$ as $g_{CAK} \sim 1/\rho^\alpha$,
we see from the steady-state mass conservation relation
$\rho \sim 1/vA \sim h(r)/vr^2$ that along  a radial flow with 
non-spherical expansion (``nse'') the line force
has the form
\beq
g_{nse}=h(r)^\alpha g_{CAK} \, .
\label{gnse-def}
\eeq
The relevant equation of motion  thus takes the same general form as 
given in eqn. (\ref{fdeom}) if we simply replace the finite-disk correction
with the non-spherical expansion, $f \rightarrow h^{\alpha}$.
With this redefinition, the dependence of the scaled acceleration 
$w'$ on $h^{\alpha}$ is thus again given by fig.~\ref{fig4}.

\begin{figure}
\begin{center}
\plotone{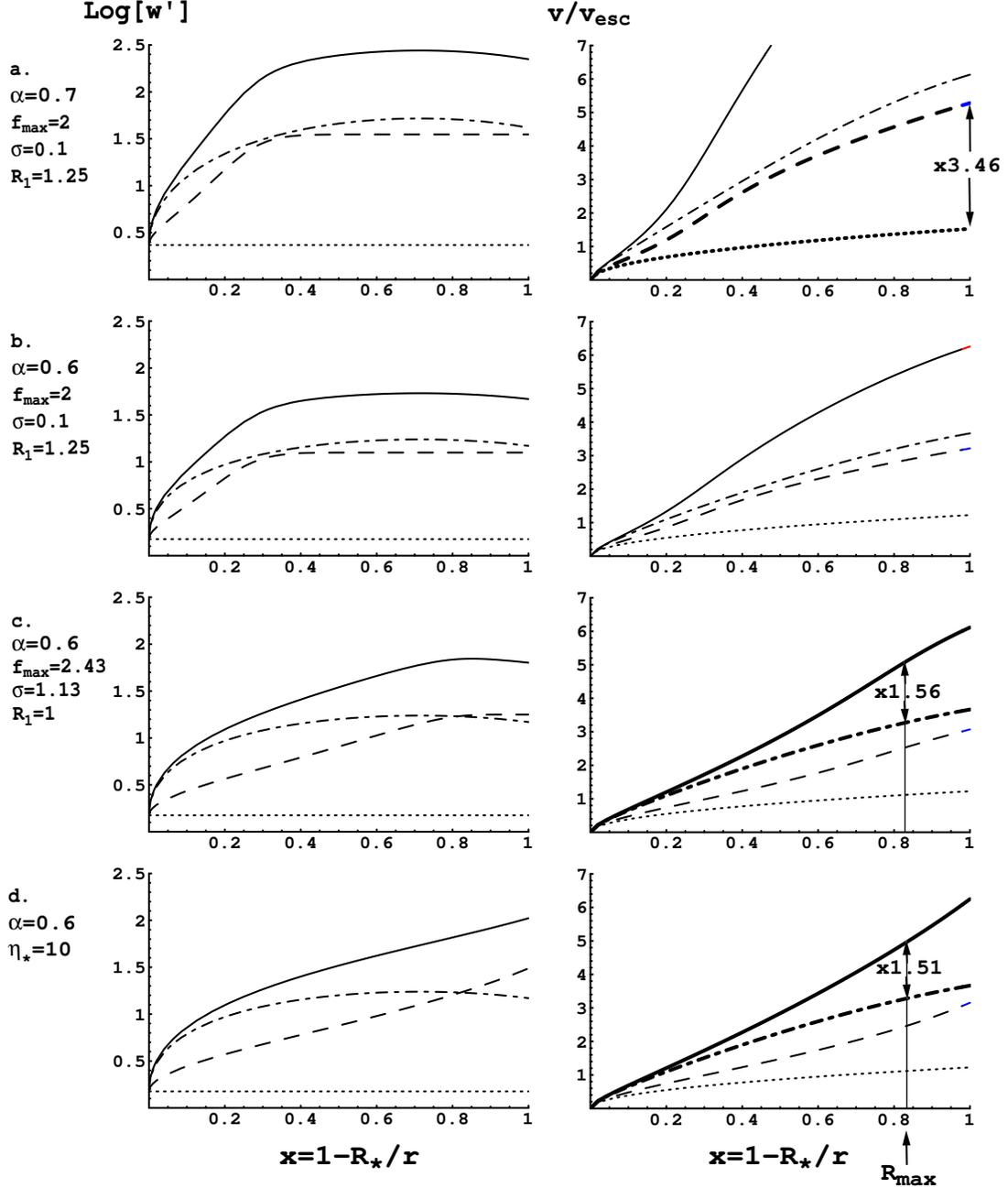}
\caption{
Results for various 1D flow models, showing 
the log of scaled acceleration $w'$ (left) and the
ratio of flow speed to escape speed $v/v_{esc}$ (right), 
plotted vs. scaled inverse radius  $x \equiv 1-R_{\ast}/r$.
The dotted and dot-dashed curves represent respectively 
the  point-star and finite-disk solutions for spherical expansion, 
while the dashed and solid curves respresent the corresponding 
non-spherical solutions.
The panels a-d represent various models, with key parameters as 
labeled on the left.
The top three models (a-c) use the Kopp-Holzer scaling in eqn. 
(\ref{fkhdef}), while the bottom model uses the scaling of eqns. 
(\ref{huo-def}) and (\ref{rcstd}), with 
$\eta_{\ast} = 10$.
}
\label{fig6}
\end{center}
\end{figure}

Fig.~\ref{fig6}a
illustrates results for this case, showing the spatial variation (in 
terms of the inverse radius coordinate $x$) of $\log w'$ (left)
and wind speed ratio $\sqrt{w} = v/v_{esc}$ (right).
(The latter is obtained by straightforward numerical integration of 
the acceleration $w'$ over $x$.)
In both panels, the dotted and dashed curves represent respectively 
the  point-star and finite-disk solutions for spherical expansion, 
while the dot-dash and solid curves respresent the corresponding 
non-spherical solutions for this standard M-88 case with $\alpha=0.7$
and $h_{max}=2$.
Note that both the acceleration and velocity laws in this point-star, 
non-spherical expansion case (dot-dash curve) agree roughly to what is found in 
the corresponding spherical, finite-disk model (dashed curve).
In this sense, the M-88 results for this $h_{max}$ can be roughly thought 
of as simply substituting the non-spherical expansion for the 
finite-disk correction.

For this case with $h_{max}=2$ we find in particular a terminal speed ratio 
$v_\infty/v_{esc} = 5.2$,  which, for the M-88 assumed stellar parameters 
that set the effective  escape speed $v_{esc}=974$~km/s, yields a terminal speed
$v_\infty \approx 5145$~km/s, or slightly above the value $v_\infty = 4799$~km/s
quoted in table 1 of M-88.
(The  modest difference is due to the neglect here gas pressure terms 
that were included in M-88's more elaborate analysis; see Appendix A 
and B.)
When compared to the terminal speed expected for point-star spherical 
expansion $v_\infty = \sqrt{\alpha/(1-\alpha)} v_{esc} = 1.53 \times 974 = 1488$
(which is slightly below the value $1512$~km/s found in M-88's 
solution with gas pressure included), we see that non-spherical expansion 
with $h_{max}=2$ leads to a speed increase of 3.46.

Table 1 summarizes the speed enhancement factors for this case, as well 
as for a series of other models to be described below.

\begin{table}[htb]
\begin{centering}
\begin{tabular}{r||cc|cc|cc}
&
\multicolumn{2}{c}{$\alpha = 0.5$} & 
\multicolumn{2}{|c}{$\alpha = 0.6$} &
\multicolumn{2}{|c}{$\alpha = 0.7$} \\ 
\multicolumn{1}{c||}{Model} & 
$r \rightarrow\infty$ & $r = 6 R_{\ast}$ &
$r \rightarrow\infty$ & $r = 6 R_{\ast}$ &
$r \rightarrow\infty$ & $r = 6 R_{\ast}$
\\ 
\hline
M-88: $h_{max}=1.5$ & 1.80 & 1.94 & 2.01 & 2.16 & 2.39 & 2.56 \\
             $=2.0$ & 2.22 & 2.39 & 2.62 & 2.81 & {\bf 3.46} & 3.69 \\ 
\hline
M-88 + $f_{d}$: $h_{max}=1.5$ & 1.26 & 1.25 & 1.37 & 1.36 & 1.60 & 1.58 \\
                       $=2.0$ & 1.47 & 1.46 & 1.71 & 1.69 & 2.22 & 2.20 \\ 
\hline
KH+UO-02: $\eta_{\ast}
                   =0.1$ & 1.10 & 1.08 & 1.13 &{\bf 1.12}& 1.21 & 1.19 \\
                  $=1.0$ & 1.26 & 1.23 & 1.38 &{\bf 1.33}& 1.65 & 1.56 \\ 
                  $=10.$ & 1.43 & 1.36 & 1.67 &{\bf 1.56}& 2.22 & 2.01 \\ 
\hline
eqn.(\ref{huo-def}): $\eta_{\ast}
                   =0.1$ & 1.13 & 1.11 & 1.19 &{\bf 1.15}& 1.29 & 1.24 \\
                  $=1.0$ & 1.26 & 1.21 & 1.38 &{\bf 1.30}& 1.64 & 1.50 \\ 
                  $=10.$ & 1.45 & 1.34 & 1.71 &{\bf 1.51}& 2.33 & 1.91 \\ 
\hline
\end{tabular}
\caption{Ratio of flow speeds between non-spherical vs. spherical 
expansion models. 
The top group of M-88 models is based on the CAK point-star model, but all 
remaining data include the finite-disk correction.
The third group uses the Kopp-Holzer non-spherical function 
(\ref{fkhdef}) with parameters set by fits from fig. 7 of UO-02 
(see also fig.~\ref{fig7} here), 
while the  bottom group uses eqn. (\ref{huo-def}) with 
$R_{c} = \eta_{\ast}^{3/8} R_{\ast}$.
The boldface values emphasize the strong speed enhancement of the M-88 point 
star model, and the more moderate values of the third and fourth model
group. Note the latter agree quite well with the MHD simulation results from 
fig.~\ref{fig1}b, which give polar speed ratios of 1.12, 1.34, and 1.54
for $\eta_{\ast}=$~0.1, 1, and 10. 
}
\end{centering}
\end{table}

A key point here is that the strong flow speed increases M-88 obtained 
from including non-spherical expansion in the point-star CAK 
model are quite analogous to the speed increases obtained by 
including a finite-disk correction.
Let us next extend our analysis to include both effects.

\subsection{Combined Effects of Non-Spherical Expansion and a Finite Disk}

For a radial flow the combined modification to the line force due to
finite disk and non-spherical expansion is given simply by the 
product of these separate\footnote{
Actually, a faster-than-radial expansion generally also implies a larger  
lateral velocity gradient, and thus alters the  finite-disk 
correction factor from its usual spherical form (\ref{fd-def}). 
In practice, we find this to be a minor correction, except very near 
the stellar surface, where it leads to a slightly increased mass 
loss rate; see Appendix B.
}
correction factors,
\beq
g_{fd,nse}=f_{d} h^{\alpha} g_{CAK} \, .
\label{gfdnse-def}
\eeq
We can thus analyze the combined dynamical effect by solving the 
equation of motion given again by eqn.(\ref{fdeom}), applying now the 
simple substitution $f \rightarrow (f_d/f_\ast)h^{\alpha}$.
With this combined rescaling, fig.~\ref{fig4} again gives the dependence 
of $\sqrt{w'}$ on the combined correction factor $f$.

The solid curves in fig.~\ref{fig6}a give the acceleration and velocity 
for the full solution of combined non-spherical expansion plus 
finite-disk.
Compared to the above point-star case, note the acceleration is
now very strong, with indeed the velocity in this case going off the scale, 
approaching an extaordinarily large terminal speed 
$v_{\infty} = 13.6 \, v_{esc} = 13,255$~km/s!
However, note that even the spherical, finite-disk model in this case 
also has a very large terminal speed of 
$v_{\infty} = 6.13 \, v_{esc} = 5971$~km/s.
As such, in {\it relative} terms of non-spherical to spherical 
enhancement, the finite-disk case implies a terminal speed factor of 
``only''2.22, which is actually significantly  less than the factor 
3.46 of the M-88 point-star assumption for this case.

Nonetheless, in {\it absolute} terms the implied terminal speeds well 
in excess of 10,000 km/s  in such a model are several times higher than 
fastest speeds (ca. 5000 km/s) inferred from blue-edges of the P-Cygni 
wind lines from ``normal'' OB stars 
(Prinja \& Howarth 1986; Howarth and Prinja 1987).
In this sense, if finite-disk effects had been included in M-88's 
original analysis, the implications of assuming hot-star magnetic 
fields would have seemed even more dramatic; 
indeed, the lack of observational evidence for such high speeds may 
well have been interpreted as strong evidence against existence of a 
dynamically significant  magnetic field 
(i.e. with confinement parameter $\eta_{\ast} > 0.1 $).

But now, with the benefit of the full MHD simulations of UO-02, we 
have a strong incentive to investigate the situation further, and 
indeed the simple formalism here makes it possible to readily
examine the effect of varying parameters, as summarized in table 1.

\subsubsection{Effect of 
More Modest
CAK Exponent $\alpha$}

The analysis in \S 3, and particularly the results in figs. \ref{fig4} 
and \ref{fig5}, suggest that a key parameter in influencing the 
wind flow speed is the CAK power-index $\alpha$.
Thus, in comparing the M-88 results with those of the UO-02 MHD 
simultions, it is important to note that the former assumed 
$\alpha=0.7$ and the latter $\alpha=0.6$.

Fig.~\ref{fig6}b thus provides results for the same Kopp-Holzer 
parameters as used in M-88 (and in fig.~\ref{fig6}a), but now with 
$\alpha=0.6$.
Note that the accelerations and speeds are all reduced.
In particular, the case with non-spherical expansion combined with 
finite-disk now has a terminal speed of $v_{\infty} = 6.28 v_{esc} = 6113$~km/s,
representing a more modest factor 
1.71 enhancement over the terminal speed
$v_{\infty} = 3.67 v_{esc} = 3570$~km/s of the corresponding 
spherical, finite-disk model.

\begin{figure}
\begin{center}
\plotone{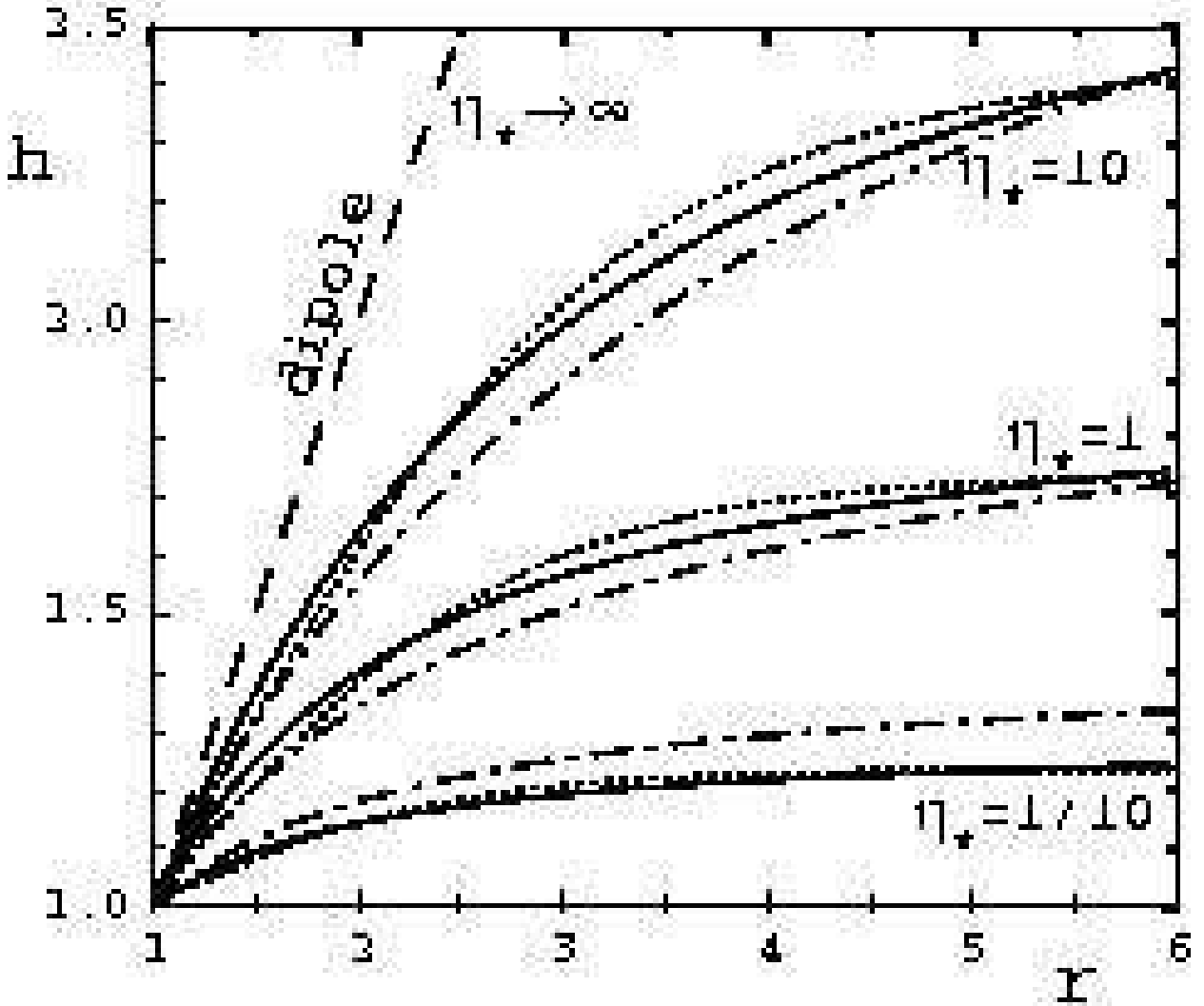}
\caption{
Non-spherical expansion factor $h(r)$ for polar flow from full MHD simulations 
of UO-02, plotted vs. radius for magnetic confinement 
parameters $\eta_{\ast} = 1/10$, $1$, and $10$ (solid curves).
The dotted curves compare the best-fit Kopp-Holzer parameterizations, 
quoted in fig. 7 of UO-02 as respectively $h_{max} =$~1.23, 1.73, and 
2.43 and $\sigma/R_{\ast}=$~0.73, 0.89, and 1.13, with $R_{1} = R_{\ast}$ 
in all cases.
The dot-dash curves here add the simple physical fit form of eqns. 
(\ref{huo-def}) and (\ref{rcstd}).
The dashed curve shows the expected result for pure dipole in limit 
of large confinement parameter, $\eta_{\ast} \rightarrow \infty$.
}
\vspace{-0.2cm}
\label{fig7}
\end{center}
\end{figure}

\subsubsection{Using Dynamically Motivated Values for the Kopp-Holzer 
Parameters}

To further facilitate comparisons with our MHD simulations results, 
let us next consider Kopp-Holzer parameter values that are adjusted to 
fit the non-spherical expansion found in these simulations.
Using eqn. (\ref{h-def}) to convert magnetic field variations into 
flow area, fig.~\ref{fig7} plots the radial variation of the polar 
expansion factor found in the MHD simulations
assuming various magnetic confinement
parameters $\eta_{\ast}$.
The dashed curve shows the expected result for pure dipole in limit 
of large confinement parameter, $\eta_{\ast} \rightarrow \infty$.
The lower, middle and upper solid curves show the MHD results for 
for $ \eta_{\ast} =$~0.1, 1, and 10.
The dotted curves plot the best-fit Kopp-Holzer parameterizations, 
quoted in fig. 7 of UO-02.

For the Kopp-Holzer fit to the strongest confinement case $\eta_{\ast} = 
10$, fig.~\ref{fig6}c shows the enhancements in acceleration and speed are 
again quite modest.
Indeed,  at the radius $R_{max} = 6 R_{\ast}$ of the outer 
boundary of the UO-02 MHD simulation, the speed of this 1D model is
$v_{6} = 5.72 v_{esc} = 5576$~km/s.
Relative to the speed of the spherical case at this radius, 
$v_{6} = 3.29 v_{esc}$~=3200~km/s, this respresents an enhancement factor 
of about $1.56$,  which is now remarkably close to the factor $\sim 
1.54$ found for the polar flow of the 2D MHD model (cf. fig.~\ref{fig1}b).

Table 1 shows similarly good agreement between this modified 
Kopp-Holzer parameterization and the MHD simulations (fig.~\ref{fig1}b)  
that show polar speed enhancements of ca. 1.12 and 1.32 for  
confinement parameters $\eta_{\ast} =$~0.1 and 1.

\subsection{A Dynamically Motivated Non-Spherical Expansion Function}

The overall variation of the MHD non-spherical expansion factors 
plotted in fig.~\ref{fig7} (and in fig. 7 of UO-02) suggests 
a simpler, more physically motivated fitting-function,
\beq
h_{UO} (r) = { r \over R_{\ast} } \, 
{ R_{\ast} + R_{c}  \over r + R_{c} } 
\, ,
\label{huo-def}
\eeq
which has just a single parameter, $R_{c}$, representing a
``magnetic confinement radius''.
This characterizes the transition between the dipole scaling $h \sim r$ that 
applies for a strong field near the surface, and the 
asymptotic monopole scaling $h \rightarrow$~constant that applies at 
large radii, where the wind has stretched the field into a nearly 
radial configuration.
Note that in the opposite limits $R_{c} \rightarrow 0$ 
or $R_{c} \rightarrow \infty$ the monopole or dipole scalings apply 
throughout the flow, as appropriate to the limits of weak 
($\eta_{\ast} \rightarrow 0$) or strong ($\eta_{\ast} \rightarrow \infty$) 
magnetic confinement.

This suggests a direct scaling of $R_{c}$ with the confinement 
parameter $\eta_{\ast}$.
In general, the ratio of magnetic to wind-kinetic energy should scale 
as $\eta (r) = $ $\eta_{\ast} (r/R_{\ast})^{2-2p}$, where $B(r) \sim 
1/r^{p}$, with $p=3$ for a dipole, and $p=2$ for a monopole.
If we then assume the confinement radius occurs where this ratio has 
some order-unity value $\eta_{c}$, we obtain a general scaling of the 
form
\beq
{ R_{c} \over R_{\ast} } \approx 
\left [ { \eta_{\ast} \over \eta_{c} } \right ]^{q} \, ,
\label{rcgen}
\eeq
where $q = 1/(2p-2)$ should be intermediate between the 
dipole ($p=3$) value $q_{d} = 1/4$ and the monopole ($p=2$) value 
$q_{m} = 1/2$.
Taking the simple case that the order-unity constant 
$\eta_{c}$ is in fact unity,
we find quite good fits for an exponent $q=3/8$, representing the 
``geometric mean'' between the dipole and monopole scalings for 
$R_{c}$,
\beq
R_{c}  \approx  \eta_{\ast}^{3/8} R_{\ast} \, .
\label{rcstd}
\eeq
The dot-dash curves in fig.~\ref{fig7} plot the radial variations 
of this simple, physically motivated form of the single-parameter 
non-spherical-expansion  function (\ref{huo-def}).
Note that the agreement with the  corresponding  MHD simulation results 
(solid) is almost as good as the best-fit, three-parameter, Kopp-Holzer form
[dotted; eqn. (\ref{fkhdef})] that has less physical justification.

Fig.~\ref{fig6}c shows the acceleration and speed when
this simple scaling is applied to the strong confinement case with 
$R_{c} = 10^{3/8} R_{\ast} = 2.37 R_{\ast}$.
At the simulation outer radius $R_{max} = 6 R_{\ast}$, the 
speed ehancement of this non-spherical, finite-disk model is now
a factor 
of about $1.51$,  which is again remarkably close to the factor $\sim 
1.54$ found for the polar flow of the 2D MHD model (cf. fig.~\ref{fig1}b).
Table 1 shows similarly good agreement between this simpler 
non-spherical enhancement factor and the  MHD polar speed 
enhancements for the other confinement parameter cases.

\section{Mass Loss for Tilted Flow}

In addition to modifying the flow divergence, a magnetic field can 
also alter a flow's direction. 
In particular, for a strong field the initial flow off the stellar 
surface occurs at an angle tilted to the local radial direction.
In this section we examine how such a base flow tilt should influence 
the local mass flux.


Since the mass loss is set very near the stellar surface, where combined factor 
from non-spherical divergence and finite-disk correction has its 
minimum, let us ignore both sphericity and magnetic divergence effects and 
 consider the field and flow to be along a {\it fixed} direction 
${\bf \hat{s}}$, assumed to be tilted by an angle $\theta_{B}$ with 
respect to the local vertical direction, defined by unit vector
${\bf \hat{z}}$. 
We then have 
$\mu_{B} \equiv \cos \theta_{B} = {\bf \hat{s}} \cdot {\bf \hat{z}} $, 
and in terms of the local horizontal direction ${\bf \hat{x}}$, we can 
also write 
$ {\bf \hat{s}} = \mu_{B} \, {\bf \hat{z}}+ \sqrt{1-\mu_{B}^2}\, {\bf 
\hat{x}}$.
 
For our usual simplification of negligible gas pressure, 
the steady-state equation of motion along the fixed field direction 
${\bf \hat{s}}$ becomes
[cf. eqn. (\ref{eom})]
\beq
({\bf v \cdot \nabla})({\bf \hat{s} \cdot v }) =
- g_{\ast}
(  {\bf  \hat{s} \cdot \hat{z}})
\,
+ {\bf \hat{s}} \cdot {\bf g}_{CAK} \, ,
\label {eom-tilt}
\eeq
where on the right-hand-side ${\bf g}_{CAK}$ represents the vector radiative 
force, and we have used the fact that surface gravity $g_{\ast} 
\equiv (1-\Gamma)GM/R_{\ast}^{2}$ is along the local vertical ${\bf 
\hat{z}}$.
On the left-hand-side the advective derivative of velocity along 
${\bf \hat{s}}$ can be written as
\beqa
\nonumber
({ \bf v \cdot \nabla})({\bf \hat{s} \cdot v})&=&
v_s({\bf \hat{s} \cdot \nabla})( v_s)
\\
\nonumber
&=&\mu_{B} \, v_s {\partial v_s \over \partial z} +\sqrt{1-\mu_{B}^2} 
v_s {\partial v_s \over \partial x}
\\
&=& \mu_{B} \, v_s {\partial v_s \over \partial z}\, ,
\label {vsdvsbdr}
\eeqa
where the last expression assumes that there are no horizontal 
variations in velocity.

For the line-driving, let us again assume a point-source 
approximation with the 
radiation confined to a pencil beam along the local vertical ${\bf 
\hat{z}}$.
In terms of an associated surface flux $F_{\ast} \equiv L/4 \pi 
R_{\ast}^{2}$,
the projected CAK line-force term then becomes 
[cf. eqn.(\ref{gcak})]
\beqa
\nonumber
{\bf \hat{s}} \cdot {\bf g}_{CAK}  &=& 
\frac {1}{1-\alpha} \,
\frac {\bar {Q} \kappa_e F_{\ast} }{ c} \,
(  {\bf  \hat{s} \cdot \hat{z}})
 {\left[ \frac { {\bf \hat{z} \cdot  \nabla}
[{\bf \hat{z} \cdot  v}]}{\bar {Q}
\kappa_e \rho c} \right]}^{\alpha} \,
\\
\nonumber
&=&
\frac {\mu_{B}}{1-\alpha} \,
\frac {\bar {Q} \kappa_e F_{\ast} }{  c} \,
\left[ \frac {\mu_{B} \partial v_s / \partial z}
{\bar {Q} \kappa_e  \rho c} \right]^{\alpha}\, 
\\
\nonumber
&=&
\frac {\mu_{B}}{1-\alpha} \,
\frac {\bar {Q} \kappa_e F_{\ast} }{ c} \,
\left[ \frac {\mu_{B} v_{s} \partial v_s / \partial z}
{\bar {Q} \kappa_e  \mdot_{s} c} \right]^{\alpha}\, 
\\
&=&
\frac {\mu_{B}}{1-\alpha} \,
\frac {\bar {Q} \kappa_e F_{\ast} }{ c} \,
\left[ \frac {\mu_{B}^{2} v_{s} \partial v_s / \partial z}
{\bar {Q} \kappa_e  \mdot_{r} c} \right]^{\alpha}\, ,
\label {gcak-tilt}
\eeqa
where the third equality uses mass conservation along the tilted flow 
to eliminate the density $\rho$ in favor of the mass flux in
the tilted direction, $\mdot_s \equiv \rho v_s$.
The final equality then uses the fact that the associated {\it 
radial} 
mass flux component $\mdot_{r}$ 
(which sets the associated mass loss 
through the spherical surface)
is reduced by another factor $\mu_{B}$, i.e. $\mdot_{r} = \mu_{B} \mdot_{s}$.
After dividing through by $\mu_{B}$ and making the modified definition
[cf. eqn.(\ref{wp-def})],
\beq
w' \equiv {v_s dv_s/dz \over  g_{\ast}} \, ,
\eeq
the momentum equation (\ref {eom-tilt}) can again be recast in the familiar 
form [cf. eqn. (\ref{cakeom})],
 \beq
w'=-1+C \mu_{B}^{2 \alpha} (w')^\alpha \, ,
\label{cakeom-tilt}
\eeq
where $C$ is defined in  eqn. (\ref{cdef}), and is related to the 
maximal
value of the radial mass flux.
Equation (\ref{cakeom-tilt}) is mathematically 
identical to eqn. (\ref {cakeom}) under the substitution 
$C \rightarrow C\, \mu_{B}^{2\alpha}$.
As such, following the same procedure as in \S 2.1, we find the 
surface mass flux  scales as
\beq
\mdot_{r}= \mu_{B}^{2} \mdot_{CAK}
\eeq
where $\mdot_{CAK} \equiv \Mdot_{CAK} /4 \pi R_{\ast}^{2}$ is the 
analogous surface mass flux  in the CAK model for spherical 
mass loss.

The $\mu_{B}^{2}$ reduction of the radial surface mass flux predicted by 
this simplified planar, 1D flow analysis is actually in quite good 
agreement with 
results of our full 2D MHD simulations (Ud-Doula 2002; UO-02). 
Indeed, the solid curves of fig.~\ref{fig1}a above show that the 
mass flux scaled by $\mu_{B}^{2}$ is roughly constant in latitude, and 
equal to the mass flux in the  spherical case.
Physically, one factor of $\mu_{B}$ can be attributed simply to geometric 
projection of the tilted flow onto the vertical normal to the star's 
spherical surface, while the other can be seen to result from the 
tilted flow acceleration having less effect in desaturating the 
optically thick absorption of the radially streaming radiation.

This lower initial mass flux combined with the faster-than-radial 
area expansion leads to a lower density, which through the CAK 
line-force leads to a greater acceleration and thus a faster terminal 
speed.
In cases with a large magnetic confinement parameter (i.e. 
$\eta_{\ast} = 10$), the simulations thus tend to show the greatest 
terminal speeds at latitudes just away from the magnetic equator, 
which originate from the open field lines that are furthest from the 
magnetic  pole, and for which the initial flow tilt at the surface wind base 
is greatest. (See fig.~\ref{fig1}b here and right panel of fig. 8 in UO-02).

\section{Discussion}

The mass flux and speed variations arising in such magnetic models of 
line-driven winds provide a potentially attractive mechanism for 
explaining the extensive wind structure and variability commonly 
inferred from observations of UV P-Cygni lines from OB stars
(e.g., Kaper et al. 1996; Prinja et al. 1998).
If the models here were extended to include even a modest stellar 
rotation along an axis that is {\it tilted} relative to the magnetic 
dipole axis, then the speed variations with magnetic latitude would 
naturally lead to Co-rotating Interaction Regions (CIRs), much as occur between 
faster and slower streams in the solar wind.
Following initial suggestions by Mullan (1984), such CIRs now represent 
a popular general paradigm for explaining the commonly observed 
``Discrete Absorption Components'' (DACs) in P-Cygni absorption troughs.
But in previous hydrodynamical simulations by Cranmer and Owocki (1996), 
the speed variations leading to CIR formation had to be induced 
somewhat artificially, by assuming a modulation of the base wind 
driving, as might result from, e.g., bright or dark spots on the stellar 
surface.
Since the analyses here (and the UO-02 MHD simulations) show that even 
modest magnetic confinement  parameters can lead to quite substantial 
latitudinal variations in flow speed, a tilted dipole field seems a 
much more likely mechanism for inducing such wind CIRs.

Note that such CIRs would also tend to limit the highest speeds 
occuring from the most rapidly diverging flow, simply from the 
truncation by the interaction with the slower speed flow.
As such the cases with the highest overall flow speeds would be those 
with a strong, {\it rotation-aligned} dipole, for which the high speed 
polar wind wouldn't be brought into interaction with the slower, 
equatorial wind.
Such cases may be relatively rare, but are possible.
The flow speeds would be quite high, perhaps up to twice that of a 
spherical wind.
But the analysis here shows that they needn't be absurdly high, at 
least for moderate values of the CAK power index, i.e. $\alpha \approx 
0.6$.

\section{Concluding Summary}

This paper analyzes the effects of the flow tilt and non-spherical
expansion associated with magnetic channeling on the mass flux and 
flow speeds of a line-driven stellar wind.
Our main results are summarized as follows:
\begin {enumerate}
\item
A faster-than-radial expansion leads to an enhancement in wind speed, 
but the relative corrections are typically about 50\% when compared to 
a finite-disk-corrected model with a moderate CAK power-index 
$\alpha=0.6$, much less than the factor three or more inferred by 
M-88 in their analysis of a point-star model with $\alpha=0.7$.
The analysis here is in good quantitative agreement with results from 
numerical MHD simulations.

\item
The non-radial expansion obtained from our numerical MHD models
differs from the heuristic form used by Kopp and Holzer (1976),
with the rapid expansion beginning right at the wind base, and extending
over a quite large radial distance. 
We propose a simpler, physically motivated fit function [eqn. 
(\ref{huo-def})] controlled by a single parameter, $R_{c}$, 
representing a characteristic radius  for wind magnetic confinement, 
and scaling as 
$R_{c} \sim  \eta_{\ast}^{3/8}$ with the magnetic confinement 
parameter $\eta_{\ast}$ ($ \equiv B_{eq}^2 R_\ast^2/{\dot M} v_\infty$).

\item
The radial mass flux at the stellar surface with a tilted magnetic 
field is reduced by the square of the cosine of the tilt-angle 
($\mu_{B}^2$) compared to a non-magnetic, spherical wind.
Physically, one factor of $\mu_{B}$ stems from the geometric 
projection of the tilted flow onto a surface normal, 
while the other arises from the reduced desaturation of absorption of 
radially streaming radiation by acceleration along this tilted flow.

\item
A perturbation analysis (in Appendix A and B) shows that the corrections 
from a small, but non-zero gas pressure scale with the ratio of sound-speed to 
escape-speed, $a/v_{esc}$.
Relative to a zero-sound-speed, finite-disk-corrected spherical wind,
typical increases in the mass loss rate are 10-20\%, with comparable 
relative decreases in the wind terminal speed.
In non-spherical models, the additional gas pressure correction is 
typically just 1-2\%.
\end{enumerate}

Overall, the results confirm that even modest magnetic fields can have a 
substantial influence in line-driven winds, with faster areal 
divergence enhancing the wind acceleration and flow speed.
But even the largest inferred enhancements still generally allow for
speeds that are within the range of observational constraints.
As such, strong magnetic fields in hot-star winds should not be precluded, 
as they might have been if previous analyses implying large speed 
enhancements had not been reduced by the more complete study here.

\def\blankline{\par\vskip \baselineskip}
\blankline
\noindent{\it Acknowledgements.}
This research was supported in part by NSF grant AST-0097983, 
NASA grants NAG5-11886 and NAG5-11095, and the NASA Space Grant College 
program at the University of Delaware.
SPO acknowledges support of a PPARC visiting fellowship and thanks J. 
Brown of the University of Glasgow and A. Willis of University College 
London for their hospitality during his sabbatical-year visits.
We thank R. Townsend and D. Cohen for helpful discussions and 
comments.

\clearpage

\appendix 

\section{ Non-Zero-Sound-Speed Correction for a Spherical, Finite-Disk 
Model}

Let us examine here the effect of a finite gas pressure associated 
with a non-zero sound speed.
In the zero-sound-speed limit considered in \S\S 3-5, the minimal value of 
the finite-disk-correction factor, $f_{d\ast} = 1/(1+\alpha)$, occuring 
near the stellar surface, $r \rightarrow R_{\ast}$, sets the critical 
conditions that determine the maximal allowed steady mass loss rate,
${\dot M}_{fd}$.
But with a finite sound speed, the reduced effective 
inertia $(1-s/w) w'$ in the base region near the sonic point shifts 
this critical location for most difficult driving slightly away from 
the surface wind base.
But then, because of the somewhat larger finite-disk factor, this 
in turn  allows a somewhat larger maximal mass loss rate.

To analyze these effects, let us  recast the equation of motion 
in the form,
\beq
F(x,w,w') \equiv (1-s/w) \, w'+ 1 
- { f \over (1+\delta m_{s})^{\alpha}}  C_{c} (w')^\alpha
- 2s/(1-x)
= 0 \,  ,
\label{fseom}
\eeq
where $s \equiv (a/v_{esc})^{2}$ is typically of order $10^{-3}$, and
\beq
\delta m_{s} \equiv { {\dot M}_{fd,fs} \over {\dot M}_{fd} } - 1
\label{mdef}
\eeq
is the fractional correction of the
mass loss rate due to this small, but non-zero $s$.
The new critical point where this mass loss is fixed now occurs somewhat away 
from the surface, $x_{c} > 0$,
but in the usual cases that $s \ll 1$, we expect both $\delta m_{s} \ll 1$
and $x_{c} \ll 1$.

In the narrow region from the sonic point to this critical point, the 
driving is dominated by the need to overcome gravity.
This implies an outward acceleration that scales as the inward gravity, 
namely $w'(x) \approx w'_{c} =  \alpha /(1-\alpha)$, 
which in turn implies that the velocity in this initial region 
nearly follows the standard CAK (``beta=1/2'') solution,
$w (x) \approx w'_{c} x$.
Using this, we obtain a spatially explicit form for the finite-disk 
factor, which upon expansion to first order in $x  \le x_{c} \ll 1$ becomes
\beq
f(x) \approx 1 + 4x \, .
\label{fxox}
\eeq
We then find the equation of motion (\ref{fseom}) takes the 
spatially explicit form
\beq
F(x) =   {(1-\alpha) s \over x}   + 4 x  - \alpha \, \delta m_{s}  + 
2s = 0 \, ,
\label{fsxeom}
\eeq
where we have also expanded $1/(1+\delta m_{s})^{\alpha}$ and kept terms to 
first order in the smallness parameter $s$.
The critical point is located where this function has a spatial 
minimum, 
\beq
0 = \left [ {dF \over dx }\right ]_{c} = 4 - {(1-\alpha) s \over x_{c}^{2}} \, , 
\label{fsxmin}
\eeq
which solves to
\beq
x_{c} \approx 
\sqrt{(1-\alpha)s}/2
\, .
\label{xcapprox}
\eeq
Applying this in eqn. (\ref{fsxeom}) and keeping only the leading term of 
order $\sqrt{s}$,  we find the mass loss correction is given by
\beq
\delta m_{s} \approx { 4\sqrt{1-\alpha} \over \alpha} \, {a \over v_{esc}}
\, .
\label{dmsapprox}
\eeq

Beyond this critical point, the increase in the finite-disk factor 
means that solutions to the equation of motion quickly approach 
their large gradient limit, 
$w'^{1-\alpha} \approx f C^{c}/(1+\delta m_{s} )^{\alpha}$.
(See dashed curves in fig.~\ref{fig4}.)
Since the wind terminal speed scales as $v_{\infty} \sim \sqrt{w'}$, 
this implies that the fractional speed corrections from a finite sound speed 
should scale as 
\beq
\delta v_{\infty,s} 
\equiv {v_{\infty,s} \over v_{\infty,s=0} } - 1
\approx (1+\delta m_{s})^{-\alpha/2(1-\alpha)} - 1
\approx { - \alpha \delta m_{s} \over 2(1-\alpha) }
\approx {-2 \over \sqrt{1-\alpha} } \, {a \over v_{esc} } \, .
\label{dvsapprox}
\eeq
For a typical case with $\alpha=1/2$ and $s \equiv (a/v_{esc})^{2}=0.001$, 
we find $x_{c} \approx 0.012$,  $\delta m_{s} \approx 0.18$, and 
$\delta v_{\infty,s} \approx -0.09 $.
For the same $s$ but 
$\alpha=0.6$, we get
$\delta m_{s} \approx 0.13$ and $\delta v_{\infty,s} \approx -0.10$,
while for
$\alpha=0.7$, we find
$\delta m_{s} \approx 0.10$ and $\delta v_{\infty,s} \approx -0.12$.
Applying these corrections to the scalings from the zero-sound-speed 
calculations in \S 3 generally gives very good agreement (to about 
a percent) with the full solutions.

It is worth noting here that, even though the 
corrections for finite gas pressure are modest, they are not, perhaps, 
as small as one might have expected. 
That is, even though the ratio of internal energy to escape 
energy is of order $s = (a/v_{esc})^{2}$, the 
corrections scale with $\sqrt{s} = a/v_{esc}$.
Moreover, this scaling is multiplied by a large coefficient, which
stems from the relatively steep increase in the leveraged 
finite disk factor -- e.g.  scaling as $f^{1/\alpha} (x) \approx 1 + 8x$ for 
$\alpha=1/2$, equivalent to an $r^{8}$ dependence on the radius $r$.
As a result, even though the reduced inertia of a mildly supersonic 
flow only leads to a relatively modest outward shift of the critical point 
from the stellar surface, the stronger finite-disk factor leads to a 
stronger radiative driving, and thus allows a larger mass loss.
Finally, that larger mass loss also implies a comparable relative reduction 
in the wind terminal speed.

\clearpage
\section{ Non-Zero-Sound-Speed Correction for a Non-Spherical, Finite-Disk 
Model}

In the case of non-spherical expansion, the analysis follows much as 
above, except that now we must include the non-spherical 
expansion factor $h(x)$.
In the region near and below the critical point $x \le x_{c} \ll 1$ this 
expands  to
\beq
h (x) \approx 1 +  h'_{o} \, x
\, ,
\label{hxox}
\eeq
where, for the form defined in eqn. (\ref{huo-def}), 
\beq
h'_{o} = { R_{c} \over R_{\ast} + R_{c} }
\approx 1/(1+\eta_{\ast}^{-3/8}) \, .
\label{hpo-def}
\eeq
This non-spherical expansion implies a stronger lateral velocity 
gradient, and for our assumed velocity law scaling 
($w \sim x$) this can be shown to modify the finite-disk correction 
expansion (\ref{fxox}) near the surface to
\beq
f_{h} (x) \approx 1 + (4 + h'_{o}) x \, .
\label{fhxox}
\eeq
Including then also the usual effect of this non-spherical expansion 
on the density, we make the replacement  $f \rightarrow f_{h} 
h^{\alpha} \approx 1+ [4 + ( 1 + \alpha) h'_{o}]  x$ and thereby find that 
equations (\ref{fsxeom}) and (\ref{fsxmin}) still apply if we simply 
make the substitution 
$ 4 \rightarrow 4 + (1+\alpha ) h'_{o}$.
The critical point thus now occurs at
\beq
x_{c} \approx 
\sqrt{(1-\alpha)s \over 4 + (1 +\alpha ) h'_{o} }
\, ,
\label{xchapprox}
\eeq
while the mass loss correction is now given by
\beq
\delta m_{s} \approx { 2 \over \alpha}
\sqrt{
(1-\alpha)
[4 + (1 + \alpha ) h'_{o}]
}  \, 
{a \over v_{esc}}
\, ,
\label{dmshapprox}
\eeq
and the speed correction by
\beq
\delta v_{\infty,s} \approx 
{ - 4 \over 
\sqrt{ (1-\alpha ) [ 4 + (1 + \alpha ) h'_{o} ] }
} \, 
{a \over v_{esc} } \, .
\label{dvshapprox}
\eeq
For the weak, moderate, and strong magnetic confinement cases 
($\eta_{\ast} =$~0.1, 1, and 10) we find 
$h'_{o} \approx $~0.3, 0.5, 0.7.
For the typical $s$ of order $10^{-3}$, we find the {\it extra} changes 
from non-spherical divergence range from one to two percent 
in both the mass loss rate and flow speed.
In particular, applying the parameters $s=0.0014$ and $\alpha=0.6$ 
used in the UO-02 MHD simulations, we estimate extra mass loss 
enhancements for the weak, moderate, and strong-field cases to be 
respectively 0.9\%, 1.5\%, and 2.1\%.
These values are roughly consistent with 
the polar mass fluxes plotted in fig.~\ref{fig1}a.

\end{document}